\documentclass{article}
\makeatother
\usepackage{spconf,amsmath,graphicx}
\usepackage{amsmath,graphicx}
\usepackage{amsbsy}
\usepackage{epsfig}
\usepackage{float}
\usepackage{graphicx}
\usepackage{color}
\usepackage{url}
\usepackage{cite}
\usepackage{amssymb}
\usepackage{graphicx}
\usepackage{lipsum}
\usepackage{float}
\usepackage{cuted}
\usepackage{balance}
\usepackage{verbatim}
\usepackage{optidef}
\usepackage{soul}

\usepackage{algorithmic}
\usepackage{algorithm2e}
\usepackage{subcaption}

\usepackage[dvipsnames]{xcolor}
\definecolor{mygreen}{rgb}{0, 0.308, 0}
\newcommand{\FP}[1]{\textcolor{black}{#1}}
\newcommand{\REB}[1]{\textcolor{black}{#1}}

\usepackage{acro} 
\DeclareAcronym{awgn}{
	short = AWGN,
	long  = additive white Gaussian noise
}
\DeclareAcronym{DDPM}{
	short = DDPM,
	long  = Denoising Diffusion Probabilistic Models
}
\DeclareAcronym{BPP}{
	short = BPP,
	long  = Bits Per Pixel
}
\DeclareAcronym{BPPF}{
	short = BPPF,
	long  = Bits Per Pixel per Frame
}
\DeclareAcronym{GAN}{
	short = GAN,
	long  = Generative Adversarial Networks 
}
\DeclareAcronym{SRR}{
	short = SRR,
	long  = Semantic Relevant Residual
}

\DeclareAcronym{SSM}{
	short = SSM,
	long  = Semantic Segmentation Map
}

\DeclareAcronym{Our_model}{
	short = SCSRDM,
	long  = Semantic-Conditioned Super-Resolution Diffusion Model
}

\DeclareAcronym{Our_model_SPIC}{
	short = SPIC,
	long  = Semantic-Preserving Image Coding based on Conditional Diffusion Models
}

\DeclareAcronym{DSSLIC}{
	short = DSSLIC,
	long  = Deep Semantic Segmentation based Learned Image Compression
}
\DeclareAcronym{SOTA}{
	short = SOTA,
	long  = state-of-the-art
}

\DeclareAcronym{SR}{
	short = SR,
	long  = Super-Resolution
}

\DeclareAcronym{mIoU}{
	short = mIoU,
	long  = mean Intersection over Union
}

\DeclareAcronym{FID}{
	short = FID,
	long  = Frechet Inception Distance
}

\ninept

\title{Semantic-preserving image coding based on Conditional Diffusion models}
\name{Francesco Pezone{\small{$^{1,4}$}}, Osman Musa{\small{$^{3}$}}, 
 Giuseppe Caire{\small{$^{4}$}}, Sergio Barbarossa{\small{$^{2}$}}\vspace{-.2cm}}
\address{\normalsize{$^{1}$DIAG Department, $^{2}$DIET Department, Sapienza University of Rome} \\
\normalsize{$^{3}$BIFOLD, $^{4}$Communications and Information Theory Group, Technische Universit{\"a}t Berlin}\\
\normalsize{{E-mail: \{francesco.pezone, sergio.barbarossa\}@uniroma1.it, 
\{osman.musa, caire\}@tu-berlin.de}}\thanks{Caire's and Musa's work was partially funded by the German Ministry for Education and Research as BIFOLD – Berlin Institute for the Foundations of Learning and Data. Barbarossa's work was funded by the Italian National Recovery and Resilience Plan (NRRP) of NextGenerationEU, partnership on “Telecommunications of the Future” (PE00000001 - program RESTART and Huawei Technology France SASU, under agreement N.TC20220919044}\vspace{-0.2cm}}
\begin{document}

\maketitle
\begin{abstract}
\textcolor{black}{Semantic communication, rather than on a bit-by-bit recovery of the transmitted messages, focuses on the meaning and the goal of the communication itself. In this paper, we propose a novel semantic image coding scheme that preserves the semantic content of an image, while ensuring a good trade-off between coding rate and image quality. The proposed \ac{Our_model_SPIC} transmitter encodes a \ac{SSM} and a low-resolution version of the image to be transmitted. The receiver then reconstructs a high-resolution image using a \ac{DDPM} doubly conditioned to the \ac{SSM} and the low-resolution image. As shown by the numerical examples, compared to \ac{SOTA} approaches, the proposed \ac{Our_model_SPIC} exhibits a better balance between the conventional rate-distortion trade-off and the preservation of semantically-relevant features. Code available at \url{https://github.com/frapez1/SPIC}}
\end{abstract}
\vspace{-.1cm}

\section{Introduction and Related Work}
\begin{figure*}
\centering\includegraphics[width=1\textwidth]{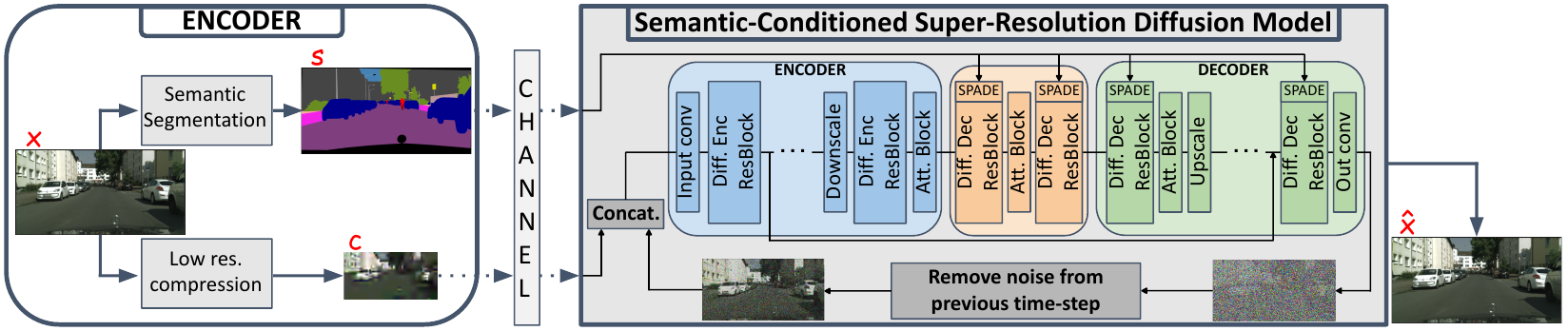}
\caption{\REB{Overview of the SPIC Architecture. The diagram illustrates our novel approach, combining a Semantic Segmentation Map ($\mathbf{s}$) and a coarse low-resolution image ($\mathbf{c}$), both compressed with classical out-of-the-shelf algorithms for efficient encoding. The reconstruction employs the proposed Semantic-Conditioned Super-Resolution Diffusion Model, leveraging both $s$ and $c$ for high-fidelity semantic-relevant image recovery even at low BPP.}}
\label{fig: schema_EncDec}
\end{figure*}
\textcolor{black}{Semantic communications is lately receiving great attention because of its potential to improve the efficiency of communication systems, focusing directly on the semantics (meaning) of the transmitted messages rather than on recovering the bits used to represent the transmitted images \cite{xuewen2022semantic, barbarossa2023semantic}. 
In semantic communication, there is no semantic error at the receiver if the reconstructed message is semantically equivalent to the transmitted one, even if the representations of the transmitted and recovered images do not coincide at the bit level. For example, in the transmission of an image captured by a web camera in an autonomous car, we might require that the reconstructed image should retain as accurately as possible the ability to detect semantically relevant objects, e.g. pedestrians, vehicles, traffic lights, etc., while providing contextually a sufficiently good trade-off between quality of the reconstructed image and the number of transmitted bits. This is just an example of combining semantics (identification of a class of meaningful objects) and the goal of communication (image reconstruction and the ability to segment the image properly at the receiver side).}
Several works have considered “semantic coding" as joint source-channel coding with semantic side information \cite{Xu2022DeepJS}.
However, since legacy protocols and network architectures are standardized according to the separation principle (OSI layers), and link layer control mechanisms do not pass erroneous data packets to the upper layers, we consider image compression at the “application layer" and do not consider transmission errors.
The fusion of semantic segmentation with image reconstruction techniques has surfaced as a potent strategy to improve the quality of image reconstruction \cite{DSSLIC, isola2017, wang2018}. By attributing semantically-relevant labels to each pixel, \ac{SSM}s represent a fundamental tool to encapsulate semantics within the image representation and can then play a key role in semantic communications.

Classical image compression techniques, such as JPEG, BPG or JPEG2000, target to achieve the best trade-off between compression ratio and image artifacts. However, this may come at the expense of semantic retention. Moreover, classical approaches can efficiently compress images without considering that some objects might be more relevant than others. 
The problem becomes even more relevant when some objects of interest have a small size, comparable to the patches used for compression. 
An example might be a pedestrian crossing the street in the distance. By applying a classical compression algorithm like JPEG2000, since the pedestrian size might occupy just a few $8 \times 8$ patches, a possible distortion in the reconstructed image might involve a small degradation of the overall image quality, but a big loss in the ability to recover crucial information like detecting the pedestrian. 

\textcolor{black}{Our goal in this work is to design compression methods able to balance high compression ratios and image quality while preserving the semantic information present in the original image.} Exploiting semantic information to guide the image reconstruction process, ensuring the preservation of crucial details of the original image, has already been considered. Recently, prominent approaches for semantic-guided image reconstruction have been built using generative models like \ac{GAN} \cite{GAN}. For example, Isola et al. \cite{isola2017} unveiled the pix2pix model, a conditional \ac{GAN} that uses a \ac{SSM} as input and outputs an image that preserves the same semantic information as the original one. Despite its visual allure, this method often overlooks the original image, resulting in a substantially different image, given the fact that the reconstruction is created starting only from the \ac{SSM}. Wang et al. \cite{wang2018} tackled the problem of reconstructing an image not close enough to the original by suggesting a conditional \ac{GAN}-centric model that integrates both the \ac{SSM} and features derived from the original image. This combination ensures better retention of the original content in the reconstructed images. Yet, these techniques often sidestep the pivotal aspect of efficient image compression since they are designed 
solely for guaranteeing an image that uses as low as possible \ac{BPP}, while optimizing a metric that does not distinguish different regions of the image, like PSNR.  Instead, we would like to tackle the problem of high-quality image reconstruction and ensure that the method can be used as a valid alternative to classical image compression algorithms, which is essential for real-world applications with bandwidth and storage limitations.

Recently,  \ac{DDPM} \cite{DDPM}, a class of generative models that match a data distribution by learning to reverse a gradual multi-step noising process, has exhibited incredible results in image synthesis \cite{DDPM_beat_GAN, kawar2022denoising, Glide}.
The authors of \cite{SISDM} improved the works of Isola and Wang et al. introducing a \ac{DDPM} model that conditions the image generation to its semantic map, hinging on the previous work \cite{SPADE}. The results obtained in these works are promising, but the regenerated images are again obtained considering the \ac{SSM} solely without taking into account the coarse image. 

\textcolor{black}{In this paper, we propose an innovative semantic image communication scheme where the transmitter encodes the \ac{SSM} losslessly, together with a low-resolution version of the image itself. The receiver uses the proposed  \ac{Our_model} to regenerate the full-resolution image. While slightly suboptimal with respect to conventional approaches, in terms of the overall rate-distortion curve, the proposed method enables a much better reconstruction and positioning of the semantically relevant objects.}  
The scheme is similar to what proposed in \cite{DSSLIC}, but with some important differences: i) our approach uses a \ac{DDPM}, as opposed to \cite{DSSLIC} that uses a GAN, because diffusion models are known for having better image synthesis capabilities \cite{DDPM_beat_GAN}; ii) differently from \cite{DSSLIC}, we do not transmit the residual error between the input and the reconstruction, to make our transmitter much simpler to implement and to limit the transmission rate; iii) instead of using a single end-to-end architecture that learns, jointly \REB{the Semantic Segmentation Map ($\mathbf{s}$) and the compressed low-resolution image ($\mathbf{c}$)}, as in  \cite{DSSLIC}, we use a modular approach that computes them separately. This simplifies the method considerably, enabling a separate control of the segmentation and compression tasks, using \ac{SOTA}  task-specific algorithms for the \ac{SSM} generation, e.g. INTERN-2.5 \cite{wang2022internimage},  and employing classical compression algorithms, e.g. BPG \cite{BPG} and FLIF \cite{FLIF}, to compress the coarse image and the \ac{SSM}. \REB{From the computational and explainability points of view, the proposed modular approach is more efficient. Exploiting off-the-shelf SOTA components, rather than training a much bigger DNN for the joint approach, allows a model with fewer parameters to train and total control over $\mathbf{s}$ and $\mathbf{c}$.
Moreover, the modular approach allows the framework to be improved easily, for example, by implementing a new \ac{SOTA} model for semantic segmentation and replacing only the semantic block without the need to retrain the whole model. } 
\section{Proposed Method}
In this section, we introduce the encoding and decoding parts of our semantic-preserving image coding based on conditional diffusion model.
\subsection{Encoder}
As illustrated in Figure \ref{fig: schema_EncDec}, the encoder is composed of two separate blocks that extract  $\mathbf{s}$ and  $\mathbf{c}$ from the input image $\mathbf{x}$. 

For the segmentation part, in this work, we used the INTERN-2.5 model \cite{wang2022internimage}, known
to have high performance in terms of semantic segmentation, but of course, as discussed before, other choices are possible.  
After generating the \ac{SSM}, we compress it with a lossless encoder since we assign high priority to the accurate reconstruction of the SSM at the receiver. More specifically, we applied the lossless compression technique FLIF, ensuring efficient encoding with an average of 0.112 \ac{BPP}. 
As far as the generation of the coarse image is concerned, we adopted different approaches. The first and simplest one is the average down-scaling operator that shrinks the image dimensions from $256 \times 512$ to $64 \times 128$. Based on top of the down-scaled version, to further compress the coarse image before transmission, we employed the BPG compression algorithm on the down-scaled image. 

\subsection{Semantically-Conditioned Super-Resolution Diffusion Model Decoder}
The Decoder takes the received SSM $\mathbf{s}$ and the coarse image $\mathbf{c}$ and sends them to the \ac{Our_model}, depicted in the right side of Figure \ref{fig: schema_EncDec}, whose goal is to reconstruct a \ac{SR} image, i.e. an image with the same dimension as the original one and similar (or even better) resolution.
\begin{figure*}[htbp]
  \centering
  
  \begin{subfigure}{0.49\textwidth}
    \centering
    \includegraphics[width=\linewidth]{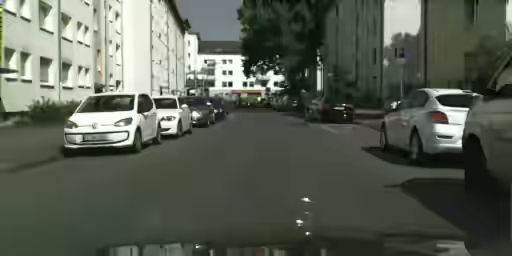}
    \caption{}
    \label{fig: comparison_bpg}
  \end{subfigure}
  \hfill
  \begin{subfigure}{0.49\textwidth}
    \centering
    \includegraphics[width=\linewidth]{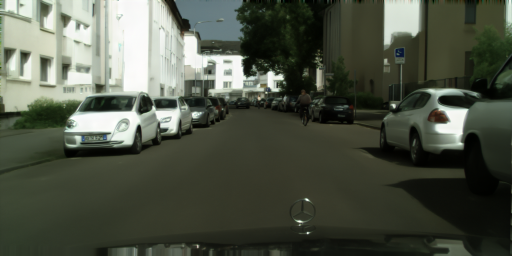}
    \caption{}
    \label{fig: comparison_OUR}
  \end{subfigure}
  
  
  \begin{subfigure}{0.49\textwidth}
    \centering
    \includegraphics[width=\linewidth]{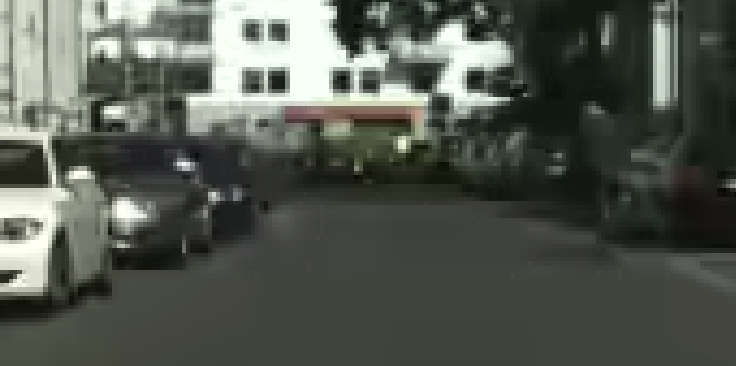}
    \caption{}
    \label{fig: comparison_bpg_detail}
  \end{subfigure}
  \hfill
  \begin{subfigure}{0.49\textwidth}
    \centering
    \includegraphics[width=\linewidth]{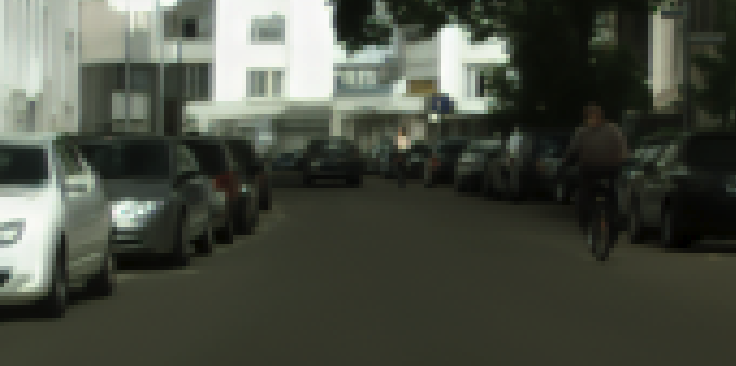}
    \caption{}
    \label{fig: comparison_OUR_detail}
  \end{subfigure}
  
  \caption{(a) Resulting image compressed with the BPG algorithm at 0.176 \ac{BPP}. (b) Reconstructed image employing our approach at 0.166 \ac{BPP}. (c) Detail of the image compressed with the BPG. (d) Detail of the image reconstructed with our approach. }
  \label{fig: BPG vs OUR}
\end{figure*}

At the model's core lies a U-net structure \cite{U-net}, encompassing three different substructures: an encoder, a central bottleneck, and a decoder. As with every \ac{DDPM}, synthesizing a singular image necessitates passing through the same U-net multiple times. At each iteration, the model inputs the previous iteration's output and the conditioning variables to predict the noise to be removed from the input image at that time step. Because of this iterative approach, \ac{DDPM}s can progressively refine the image, starting from white noise. Because of the random nature of this process, it is necessary to direct the denoising process to avoid a purely random generation disconnected from the original image. Different approaches \cite{preechakul2021diffusion, kawar2022denoising, ho2021classifierfree} can be employed to avoid this purely random generation. The two main methods are guidance and conditioning; in this work, we adopt a conditioning approach,  as it allows us to implement the dual conditioning process, which lies at the core of the proposed \ac{Our_model}, in an efficient manner.
As stated before, the idea builds on the very foundation of \ac{DDPM} and, more specifically, on the concept of \ac{SR} Diffusion Models \cite{Seper-res_refinement}, properly modified to guarantee the dual conditioning used in our strategy. 

Differently from a classical \ac{DDPM}, our  \ac{SR} Diffusion Model, during the learning phase, instead of starting the denoising process from white noise alone, concatenates the noise with a coarse image expanded to its original size. This conditioning recurs at every step during training, ensuring that the model is consistently driven from the coarse image. Specifically, during training, the model starts with a tensor of dimensions $6 \times 256 \times 512$, with the initial three (colour) channels representing white noise and the subsequent three channels containing the expanded coarse image. Throughout the training, the model undergoes $1000$ iterations to operate the transition from the white noise of the first three channels to the reconstructed image and leave the coarse image conditioning unchanged. During inference, only $20$ iterations are executed to save time (and energy), using as input always a tensor of size $6 \times 256 \times 512$, but this time substituting the noise of the first three channels with the coarse image itself. This ensures that the starting point is closely aligned with the original image.
The second conditioning is the one on the \ac{SSM}. To do so, we adapted the SPADE technique \cite{SPADE} to our model. As shown in Fig. \ref{fig: schema_EncDec}, the conditioning occurs at every ResBlock layer \cite{ResBlock} of the bottleneck and decoder subnetworks of the U-net, as also depicted in \cite{SISDM}. 
In essence, the proposed \ac{Our_model} introduces a contextual diffusion strategy, conditioned on dual inputs, achieving superior \ac{SR} outcomes, properly steered by the \ac{SSM}.

\section{Numerical Results}
\vspace{-1mm}
In this section, we delve into a comprehensive presentation of the results and advantages associated with the proposed \ac{Our_model_SPIC}. It is important to clarify that while the images utilized for these comparative analyses are sourced from the validation folder of the Cityscapes dataset, none of these images were employed during the training or validation phase.

\begin{figure}[!htbp]
\centering
\includegraphics[width=1\linewidth]{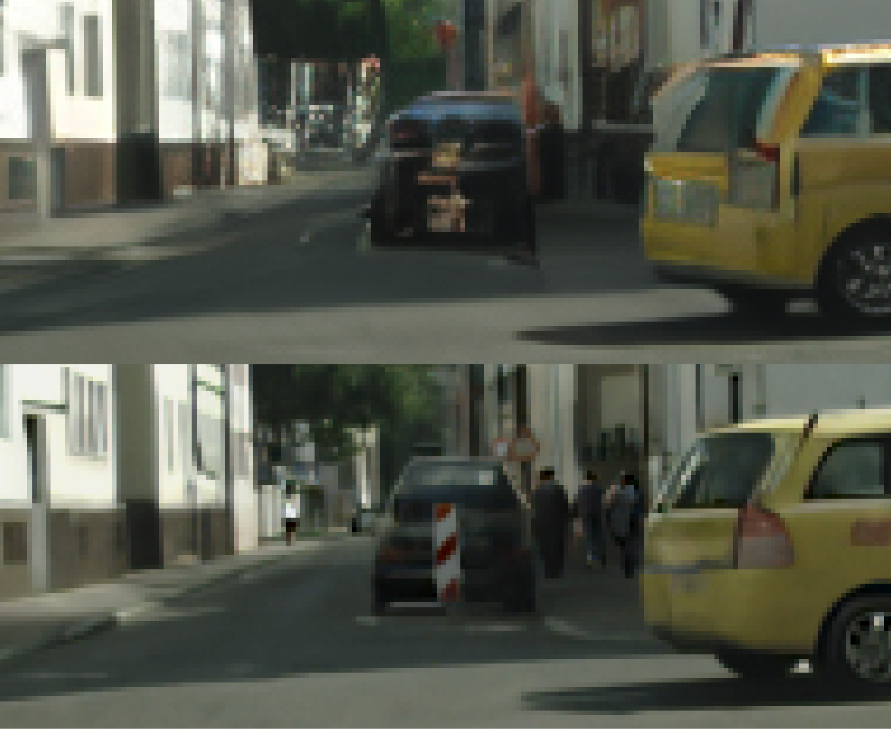}
  \caption{Detail comparison between (top) the image reconstructed with the \ac{SOTA} \ac{SR} model \cite{Super_res_CVPR} and (bottom) the image reconstructed with our model }
  \label{fig: SOTA vs OUR}
  \vspace{-2mm}
\end{figure}

As mentioned before, a paramount advantage of our model is its capability to retain semantic information while able to provide a good trade-off between the overall image quality and compression rate. Several existing compression algorithms and \ac{SR} models often reconstruct visually pleasing images. However, a closer inspection reveals a significant drawback: the degradation of semantic content, particularly evident as the size of the semantically relevant objects within the image diminishes. For larger foreground objects, most available approaches are able to detect and generate the correct semantic segmentation correctly. However, as the object size shrinks, conventional models falter, failing to accurately process the image and evaluate a precise \ac{SSM}. This aspect can be grasped by looking at Figure \ref{fig: BPG vs OUR}: on the left, we see the image reconstructed after compression with a BPG algorithm (a) and its zoom on the center part (c); on the right, we observe the image reconstructed using our approach ((b), and the corresponding detail (d). At first glance, even because of the little advantage in \ac{BPP} (0.176 vs 0.166), the image on the left looks clearer and more detailed than the one on the right. But, as soon as we zoom in, the story is completely different because our reconstruction clearly shows a person on a bicycle on the right and a pedestrian in the distance, which are not at all clear in the left image. 
As a further example, in Figure \ref{fig: SOTA vs OUR},  we compare the reconstruction capabilities of our model with the \ac{SOTA} \ac{SR} model introduced in \cite{Super_res_CVPR}.
Both models are evaluated on their ability to amplify the image size by a factor of four, transitioning from $128 \times 64$ to $512 \times 256$ pixels, without any further source compression. 
The distinguishing difference between the two approaches is that our model is conditioning the reconstruction on the \ac{SSM}. Looking at Fig. \ref{fig: SOTA vs OUR}, which reports the zoom on the central part of the reconstructed images, we can see that the resolution of our model is better and, more specifically, the three pedestrians between the two cars and the road signs are clearly visible in our case, while they are only barely observable using the SOTA 
\ac{SR} model proposed in \cite{Super_res_CVPR}.

\begin{figure}[!htbp]
\centering
\begin{subfigure}{1\linewidth}
        \centering
        \includegraphics[width=0.97\linewidth]{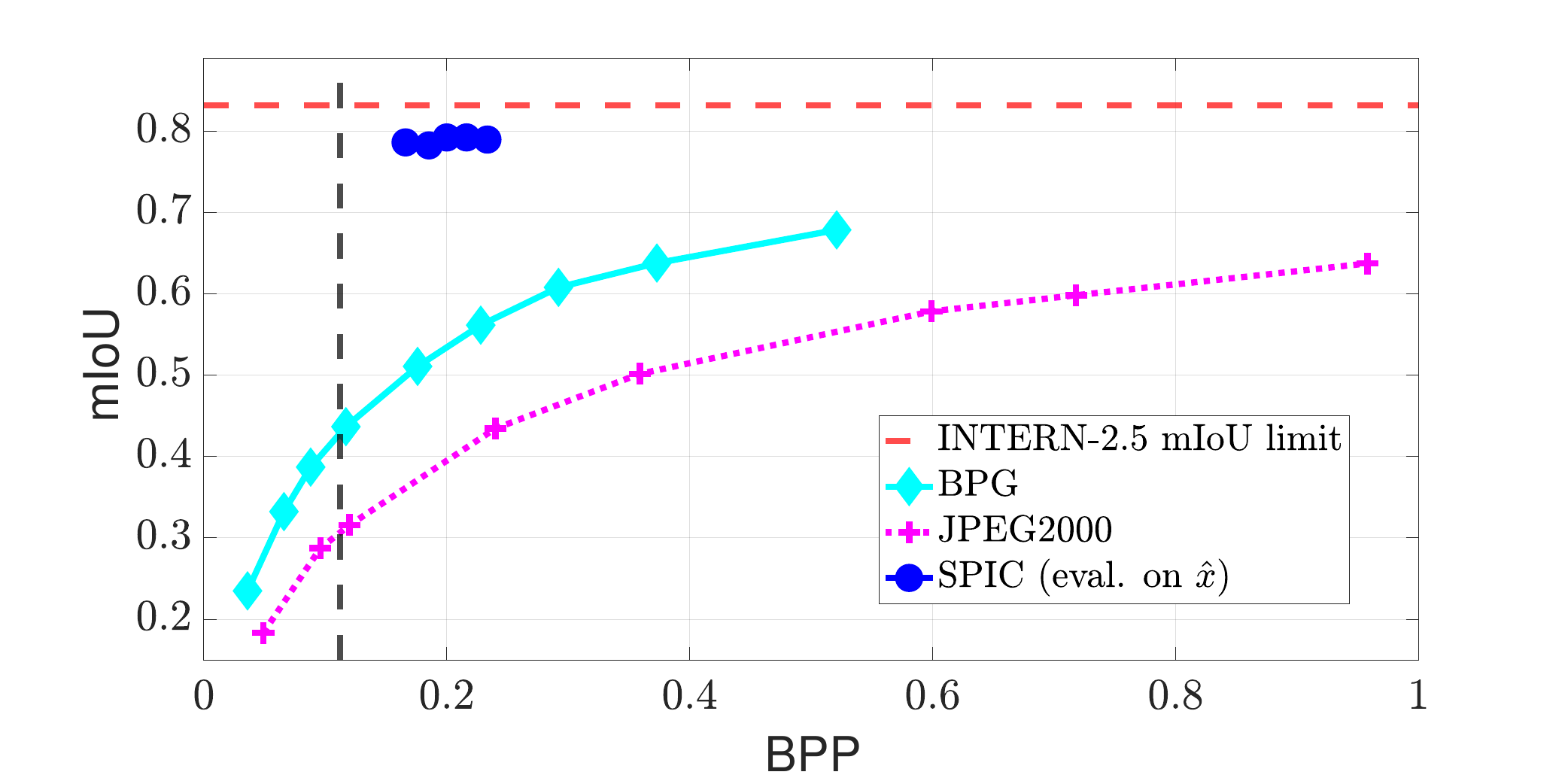}
        \caption{mIoU vs BPP}
        \label{fig: mIoU vs BPP}
    \end{subfigure}
    

    \begin{subfigure}{1\linewidth}
        \centering
        \includegraphics[width=0.97\linewidth]{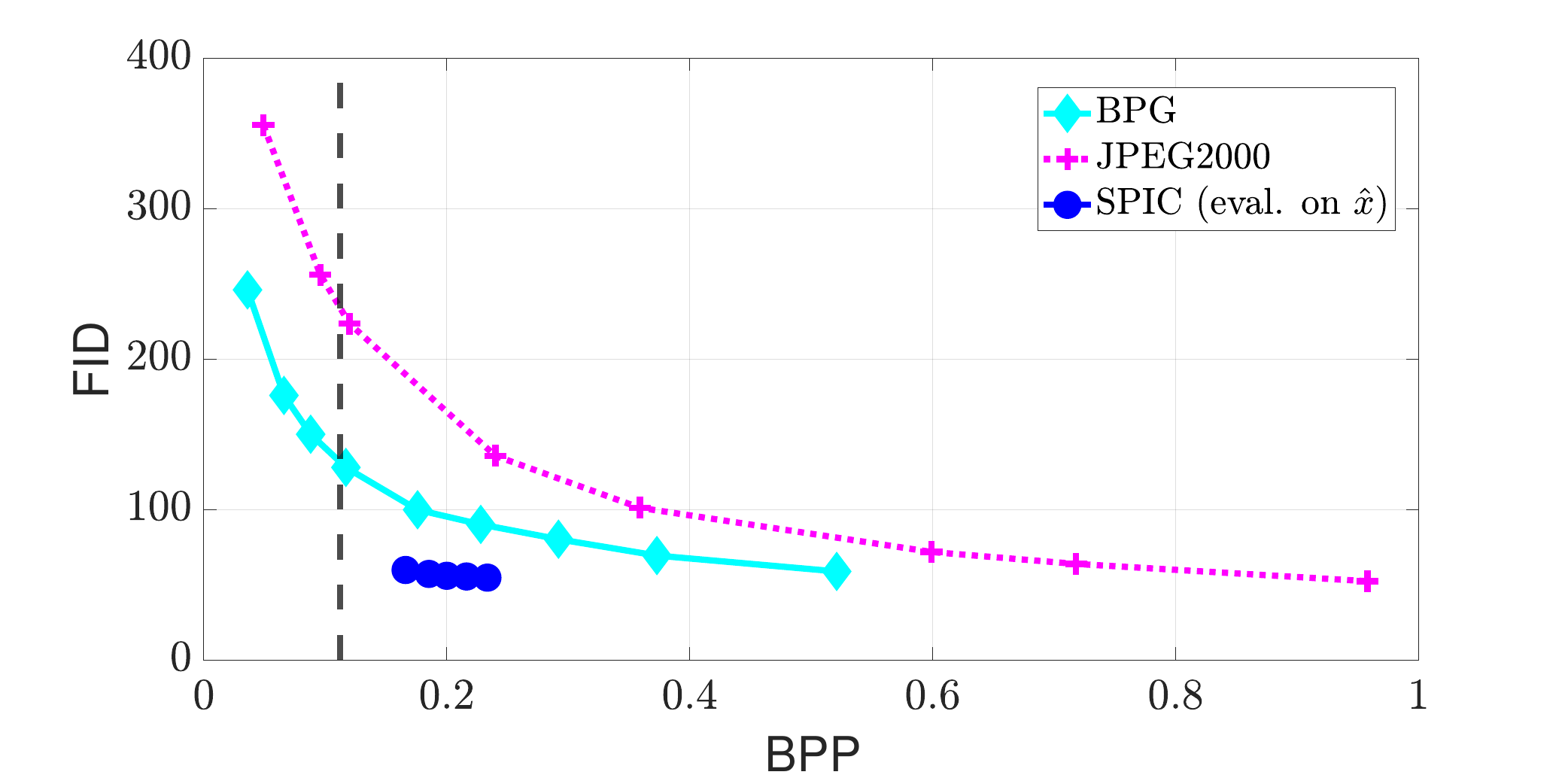}
        \caption{FID vs BPP}
        \label{fig: FID vs BPP}
    \end{subfigure}

\caption{Comparison in terms of mIoU (a) and FID (b) vs. BPP.}
  \label{fig: metrics}
\end{figure}

To compare the performance of our model with available alternatives, in terms of semantic segmentation retention, we used as a performance metric the \ac{mIoU}, a number that quantifies the degree of overlap between the ground truth and the predicted regions corresponding to the objects of interest. More specifically, given two boxes  $s_1^i$ and $s_2^i$, with $i=1, \ldots, n_c$, computed over the ground truth and the reconstructed image, where $n_c$ is the number of classes of objects of interest, the $\text{mIoU}(s_1,s_2)$ is defined as follows: 
\vspace{-2mm}
\begin{equation*}
    \text{mIoU}(s_1,s_2)  = \frac{1}{n_c} \sum_{i=1}^{n_c} \text{IoU}(s_1^i,s_2^i)= \frac{1}{n_c} \sum_{i=1}^{n_c} \frac{|s_1^i \cap s_2^i|}{|s_1^i \cup s_2^i|}
\vspace{-2mm}
\end{equation*}
In the given semantically-preserving coding scheme, the quality of the reconstructed image cannot be assessed by using conventional metrics, like PSNR, which focus only on a pixel-by-pixel reconstruction, and then fail to capture the semantic content. For this reason, since the reconstructed images are also sensitive to various types of distortion, such as blurriness, noise, and artifacts, we assess the difference between the original and reconstructed images in terms of the \ac{FID} \cite{FID}, a widely used metric in computer vision, which compares the features maps extracted by an Inception-v3 model \cite{Inception-v3}, and is expressed as follows:
\vspace{-1mm}
\begin{equation*}
    \text{FID} = \Vert \mu_{f(x)} - \mu_{f(y)} \Vert^2 + \text{Tr}(\Sigma_{f(x)} + \Sigma_{f(y)} - 2(\Sigma_{f(x)} \Sigma_{f(y)})^{1/2})
\vspace{-1mm}
\end{equation*}
with $f(\cdot)$ the output of the \textit{pool3} layer of the Inception-v3, and  $\mu$ and $\Sigma$ are the mean and covariance matrix of the 2048 feature vectors.
\vspace{-1mm}
\\
In Figure \ref{fig: metrics} (a) and (b) we report the mIoU and the FID, vs. the \ac{BPP}, used to encode the transmitted data, averaged over the whole validation dataset. 
All the \ac{SSM} are generated using the INTERN-2.5 model. In Figure \ref{fig: mIoU vs BPP}, the black dotted vertical line, positioned at 0.112 \ac{BPP}, represents the \ac{BPP} required for the lossless compression of the \ac{SSM}. The blue point represents the mIoU evaluated on the reconstructed images $\hat{x}$ obtained applying \ac{Our_model} at a \ac{BPP} given by the sum of the \ac{BPP} necessary for the lossless encoding of the \ac{SSM} and the lossy encoding of the coarse image.
The green and magenta curves represent the results achieved with BPG and JPEG2000 compression methods.
We can clearly see that both BPG and JPEG2000 exhibit worse performance than our method in terms of \ac{mIoU}. To let BPG be able to achieve \ac{mIoU} results akin to our model, the rate should be in the order of 1 \ac{BPP}.\\
Looking now at Figure \ref{fig: FID vs BPP}, we can see that while being able to retain most of the semantic segmentation information, our method can reconstruct images that have a low \ac{FID} score, outperforming both JPEG2000 and BPG.

In summary, our numerical results show that the proposed method, compared to alternative approaches,  achieves a better balance between fidelity reconstruction and ability to extract semantic features from the reconstructed image. 

\vspace{-2mm}

\vspace{-3mm}
\section{Conclusions}
In this work, we propose a novel image coding scheme, building on a doubly conditioned super-resolution diffusion model, able to better preserve the semantic content of the image than \ac{SOTA} compression algorithms and \ac{SR} methods while at the same time, having a better rate/quality trade-off when compared to the best compression methods. 
The proposed model harnesses the power of dual conditioning on a \ac{SSM} and a compressed version of the original image. The double conditioning is obtained with a modular framework that allows \ac{Our_model_SPIC} to be easily adapted to different tasks. Future investigations include the extension to semantic video coding and the incorporation of errors due to transmission over a noisy channel.

\newpage
\balance
\bibliographystyle{IEEEbib}
\bibliography{biblio}

\end{document}